\documentclass[final]{siamltex}

\usepackage{amsmath}
\usepackage{graphicx}
\usepackage[hidelinks]{hyperref}
\usepackage{cite}
\usepackage{subfig}

\begin{document}

\title{A Proximity Measure using Blink Model}

\author{
Haifeng Qian\thanks{IBM T. J. Watson Research Center, Yorktown Heights, New York ({\tt qianhaifeng@us.ibm.com}).}
\and 
Hui Wan\footnotemark[1]
\and 
Mark N. Wegman\footnotemark[1]
\and 
Luis A. Lastras\footnotemark[1]
\and 
Ruchir Puri\footnotemark[1]}

\maketitle

\begin{abstract}
This paper proposes a new graph proximity measure.
This measure is a derivative of network reliability.
By analyzing its properties and comparing it against
other proximity measures through graph examples,
we demonstrate that it is more consistent with human intuition
than competitors.
A new deterministic algorithm is developed
to approximate this measure with practical complexity.
Empirical evaluation by two link prediction benchmarks,
one in coauthorship networks and one in Wikipedia,
shows promising results.
For example, a single parameterization of this measure
achieves accuracies that are 14--35\% above the
best accuracy for each graph of all predictors reported in the 2007
Liben-Nowell and Kleinberg survey.
\end{abstract}

\pagestyle{myheadings}
\thispagestyle{plain}
\markboth{H. QIAN ET AL.}{BLINK PROXIMITY}

\section{Introduction}  \label{sec:intro}

Humans have intuitions for graph proximity.
Given a graph, certain pairs of nodes are perceived to have
stronger relation strength than others.
We know that a larger number of shorter paths indicate
greater proximity, yet a precise mathematical formulation of such perception is elusive.
Many measures in the literature 
can be viewed as quantitative
proxies of graph proximity:
shortest path, Jaccard index \cite{jaccard}, Katz \cite{katz},
personalized PageRank \cite{pr_paper},
SimRank \cite{simrank}, Adamic/Adar \cite{adamicadar}
and others \cite{potamias,blondel,fouss,koren2006,tong,yen,lao,liu,chebotarev,barabasi,leicht}.
Although they each have characteristics that suit specific applications,
they generally have varying degrees of agreement with human intuition.

This manuscript adds one more entry to the list.
This graph proximity measure is called the Blink Model and
is a derivative of network reliability.
By studying its properties and a series of graph examples,
we argue that it matches human intuition better than
many existing measures.
We develop a practical algorithm to approximately compute this measure,
and demonstrate its predicting power through empirical validations.
Some of the contents appeared in \cite{patent}.

Relational data, or graph-structured data, are ubiquitous.
Graph proximity measures, i.e., the ability to quantify relation strength,
are fundamental building blocks in many applications.
They can be used to recommend new contacts in social networks \cite{liben},
to make product recommendations based on a graph model of products and users \cite{aggarwal},
to rank web search results or documents in general \cite{pr_paper},
or to predict new facts in knowledge graphs \cite{cbmm}.
They can be used to single out anomalies by identifying implausible links.
The list of applications goes on and on.

The proposed measure is a derivative of 
terminal network reliability \cite{ball}, which has
other forms in various fields to be reviewed in Section~\ref{sec:liter}.
Network reliability has largely been ignored as a candidate measure in the
aforementioned applications.
For example, \cite{potamias} concluded that network reliability was
one of the least predictive measures.
We prove the opposite conclusion with our Blink Model measure
by including the winning measure from \cite{potamias} in both
theoretical and empirical comparisons.
The discrepancy may be due to that \cite{potamias} used a Monte-Carlo approximation
with only 50 samples which were not sufficient to reach accurate results,
and running a sufficient number of samples might have been computationally infeasible.

Exact evaluation of the Blink Model measure has the
same complexity as terminal network reliability,
which is known to be \#P-complete \cite{valiant}.
We will present a new deterministic algorithm
that approximates the measure directly with practical complexity
and thereby enables the proposed measure in applications.

To quantify the benefit of being consistent with human intuition,
we use two link prediction tasks
to compare our measure against other topological proximity measures.
The first is a replication of \cite{liben}.
A remarkable conclusion of \cite{liben} was that the best proximity measure is case dependent.
A specific parameterization of a specific measure may perform
well on one graph yet underperform significantly
on another, and there does not exist a consistent winner.
We compare against the oracle, i.e. the highest accuracy for each graph
of all predictors in \cite{liben}, and demonstrate that
a single parameterization of our measure outperforms the oracle by
14--35\% on each graph.
The second task is predicting additions of
inter-wikipage citations in Wikipedia
from April 2014 to March 2015, and again substantial accuracy advantage is shown.
Through these tasks, we also demonstrate a simple yet practical and automatic method of
training graph weighting parameters,
which can naturally incorporate application-specific domain knowledge.

This manuscript is organized as follows.
The rest of this section defines the problem, the proposed proximity measure
and several competing measures,
and reviews related fields and basic arithmetic.
Section~\ref{sec:metric} studies properties of the measure and compares
it with others through graph examples.
Section~\ref{sec:algorithm} describes the proposed approximation algorithm.
Section~\ref{sec:result} presents empirical comparisons.

\subsection{Problem statement}  \label{sec:ps}

The problem statement for a proximity measure is the following.
The input is a graph $G = \langle V,E \rangle$,
its node weights $w_V:V \rightarrow \left( 0, 1 \right]$,
and its edge weights $w_E:E \rightarrow \left( 0, 1 \right]$.
The output is, for any pair of nodes A and B in $V$, a value score(A,B).

Note that not all proximity measures consider $w_V$ and $w_E$.
Some use $w_E$ and ignore $w_V$, while some consider only topology $G$.
Although $w_V$ and $w_E$ can be from any source,
we present a simple yet practical method in Section~\ref{sec:weight}
to train a few parameters and thereby set $w_V$ and $w_E$.
It's applicable to all proximity measures and is used in Section~\ref{sec:result}.

We will focus on directed simple edges.
Undirected edges can be represented by two directed edges,
and most discussions are extendable to hyperedges.

\subsection{Definitions}  \label{sec:def}

Let us first define the proposed graph proximity measure.
Consider the input $G$ as a graph that blinks:
an edge exists with a probability equal to its weight;
a node exists with a probability equal to its weight.
Edges and nodes each blink independently.
A path is considered existent if and only if
all edges and all intermediate nodes on it exist;
note that we do not require the two end nodes to exist.
The proposed proximity measure is 
\begin{eqnarray}
s \left( \text{A},\text{B} \right) & = & - \log \left( 1-b \left( \text{A},\text{B} \right) \right)  \label{eq:metric} \\
\quad\textrm{where}\; b \left( \text{A},\text{B} \right) & = & \text{P} \left[ \text{ at least one path exists from A to B } \right] \label{eq:nr}
\end{eqnarray}
We will refer to (\ref{eq:metric}) as the Blink Model measure,
its properties and generalizations to be presented in Section~\ref{sec:metric}.
It is straightforward to see that $s$ and $b$ are monotonic functions of each other
and hence order equivalent, and the reason to choose $s$ over $b$ will be evident in Section~\ref{sec:prop}.

Next we define several competing measures.
For brevity, SimRank \cite{simrank} and commute-time \cite{fouss} are omitted
and they are compared in Section~{\ref{sec:graphs}}.

Personalized PageRank (PPR) \cite{pr_paper} with weights
considers a Markov chain that has the topology of $G$
plus a set of edges from each node to node A.
These additional edges all have transition probability of $\alpha \in \left( 0, 1 \right)$.
For each original edge $e \in E$, let X be its source node,
let $w_{\textrm{sum,X}}$ be the sum of weights of X's out-going edges,
and the transition probability of edge $e$ is $\left( 1-\alpha \right) \cdot w_E(e) / w_{\textrm{sum,X}} $.
The PPR measure $\text{score}_\text{PPR} \left( \textrm{A,B} \right)$
is defined as this Markov chain's stationary distribution on node B.
PPR does not use node weights.

The original Katz measure \cite{katz} does not use edge or node weights,
and we define a modified Katz measure:
\begin{equation} \label{eq:katz}
\text{score}_\text{Katz} \left( \textrm{A,B} \right) = \sum_{l=1}^{\infty}{ \beta^l \cdot \sum_{\textrm{length-}l\textrm{ A-to-B path }i}{p_{i,l}} }
\end{equation}
where $\beta\in \left( 0, 1 \right)$ is a parameter,
and $p_{i,l}$ is the product of edge weights
and intermediate node weights for the $i^\textrm{th}$ path with length $l$.
This measure is divergent if $\beta$ is larger than the reciprocal
of the spectral radius of the following matrix $M$:
entry $M_{i,j}$ is the product of the $i^\textrm{th}$ node weight and
the sum of edge weights from the $i^\textrm{th}$ node to the $j^\textrm{th}$ node.

The effective conductance (EC) measure is defined as
the effective conductance between two nodes by viewing edges as resistors.
It can be generalized to be a directed measure, and notable variants include
cycle-free effective conductance (CFEC) \cite{koren2006}
and EC with universal sink \cite{faloutsos,tong}.

Expected Reliable Distance (ERD) is the winning measure in \cite{potamias}.
Consider the same blinking graph as in the Blink Model
and let $D$ be the shortest-path distance from A to B,
ERD is an inverse-proximity measure:
\begin{equation} \label{eq:relidist}
\text{score}_\text{ERD} \left( \textrm{A,B} \right) = \text{E} \left[ D | D \neq \infty \right]
\end{equation}

Our implementation of Adamic/Adar \cite{adamicadar} is:
\begin{equation} \label{eq:adamicadar}
\text{score}_\text{AA} \left( \textrm{A,B} \right) = \sum_\textrm{C}{\frac{n_\textrm{A,C} \cdot n_\textrm{C,B}}{\log{d_\textrm{C,in}}+\log{d_\textrm{C,out}}}}
\end{equation}
where $n_\textrm{A,C}$ is the number of A-to-C edges,
$n_\textrm{C,B}$ is the number of C-to-B edges,
and $d_\textrm{C,in}$ and $d_\textrm{C,out}$ are the numbers of in-coming and out-going edges of node C.

\subsection{Related work}  \label{sec:liter}

This section briefly reviews fields related to the proposed measure.

Network reliability is an area in operations research
which focuses on evaluating two-terminal reliability, i.e. (\ref{eq:nr}),
all-terminal reliability and other similar quantities.
Target applications were
assessing the reliability of ARPANET, tactical radio networks, etc.
Complexity of exact evaluation of (\ref{eq:nr}) was proved to
be \#P-complete \cite{valiant}.
Fast exact methods were developed for special topologies \cite{satyanarayana,politof}.
Deterministic methods were developed for evaluating bounds \cite{ball,brecht}.
Monte Carlo methods were developed for estimation \cite{vanslyke,fishman,karger},
and were considered the choice for larger graphs with general topologies \cite{ball}.
Most of these methods have poor scalability to large graphs.
In a work \cite{hardy} in 2007,
which was a modern implementation based on binary decision diagrams (BDD),
the largest square-2D-grid benchmark had only 144 nodes.
In another work \cite{ramirez} in 2005, which used a Monte Carlo method,
the largest benchmark had only 11 nodes.

Our blinking graph definition belongs to
the category of random graphs \cite{gilbert,erdos}.
It is a generalization of the Edgar Gilbert model
by having a different probability for each edge
(zero probabilities remove edges from the Edgar Gilbert model).
In particular, the branch of percolation theory \cite{kesten,bollobas}
and works on uncertain graphs \cite{potamias,jin,zou,yuan}
are closely related to our study.

Network influence studies the spread of influence through social networks.
A popular model, called independent cascade model \cite{kempe03},
considers the same blinking graph with all node weights being 1,
and the influence $\sigma\left( S \right)$ is, given a set of starting nodes $S$,
the expected number of reachable nodes from $S$.
In the case of $S=\{A\}$,
$\sigma\left( S \right)$ is equal to the sum of (\ref{eq:nr}) over all node B's.
The goal of optimization is to select $S$
to maximize $\sigma\left( S \right)$.
Since the quantity of interest is sum of (\ref{eq:nr}) values,
it is easier to compute than individual (\ref{eq:nr}) values.
For example, many fewer Monte Carlo samples are needed to reach the same
relative error.
Methods have been developed to quickly estimate $\sigma\left( S \right)$ \cite{chen,wang},
e.g., \cite{wang} uses the largest-probability single path to each node
as a surrogate for an individual (\ref{eq:nr}) value.
Although these methods showed fidelity at the $\sigma\left( S \right)$ level,
they incur too much error for our purpose.

\subsection{Basic arithmetic}  \label{sec:basic}

For clarity of presentation and without loss of generality\footnote{Blinking
graphs with edge weights alone, setting all node weights to 1,
are equally expressive.
A node weight can be expressed as an edge weight by splitting a node
into two nodes, one being sink of in-coming edges, one being source of out-going edges,
and adding an auxiliary edge between the two, with edge weight
equal to the original node weight \cite{ball}.},
this section assumes all node weights being 1.

Exact evaluation of the Blink Model can be as follows.
Enumerate all subgraphs of $G$, each of which is a state of the blinking graph
and has a probability that is equal to the product of $w_E(e)$ for edges $e$ that exist
and $1-w_E(e)$ for edges $e$ that do not.
(\ref{eq:nr}) is the sum of probabilities of subgraphs where a path exists from A to B,
and (\ref{eq:metric}) gets calculated accordingly.
This has impractical complexity.

Monte Carlo evaluation of the Blink Model measure can be as follows.
Each sample traverses the subgraph reachable from A in one instance of the blinking graph.
(\ref{eq:nr}) is approximated by the fraction of samples that reach B,
and (\ref{eq:metric}) gets approximated accordingly.
This can be expensive.
In Table~\ref{tbl:toy}, we demonstrate that
at least tens of thousands of samples are needed to reliably discern different pairs of nodes.
Yet in Section~\ref{sec:wikipedia} for a graph that represents Wikipedia citation network,
practical run time allows only 100 samples.

If edges $e_1=(\textrm{X},\textrm{Y})$ and $e_2=(\textrm{Y},\textrm{Z})$ are the only edges to/from node Y,
they can be replaced by a single edge from X to Z with weight $w_E(e_1)\cdot w_E(e_2)$,
without altering the Blink Model measure for any pair of nodes.

Two parallel edges $e_1$ and $e_2$ can be replaced by a single edge with weight $1-(1-w_E(e_1))\cdot(1-w_E(e_2))$,
without altering the Blink Model measure for any pair of nodes.
$x$ parallel edges with weight $w$ is equivalent to a single edge with weight $1-(1-w)^x$.
In other words, an edge with weight $1-(1-w)^x$ is $x$ times as strong as an edge with weight $w$.

\section{The Measure} \label{sec:metric}

This section starts with two important properties of the proposed proximity measure,
followed by its generalizations and variations.
We then use a series of graph studies to demonstrate that the proposed measure
is more consistent with human intuition than competitors,
and finally discuss setting edge and node weights in applications.

\subsection{Properties} \label{sec:prop}

Let us begin with two properties of (\ref{eq:metric}), \emph{additivity} and \emph{monotonicity},
which are important in the coming sections.

\emph{Additivity}.
Let $G_1 = \langle V_1,E_1 \rangle$ and $G_2 = \langle V_2,E_2 \rangle$ be two graphs
such that $V_1 \cap V_2 = \left\{ \textrm{A,B} \right\} $, $E_1 \cap E_2 = \emptyset $.
Let $G_3 = \langle V_1 \cup V_2 ,E_1 \cup E_2 \rangle$ be a combined graph that has the same
node and edge weights as in $G_1$ and $G_2$.
Let $s_{G_1} \left( \text{A},\text{B} \right)$, $s_{G_2} \left( \text{A},\text{B} \right)$ and
$s_{G_3} \left( \text{A},\text{B} \right)$ be the measure value (\ref{eq:metric}) in these
three graphs respectively.
Then this condition holds:
\begin{equation} \label{eq:add}
s_{G_3} \left( \text{A},\text{B} \right) = s_{G_1} \left( \text{A},\text{B} \right) + s_{G_2} \left( \text{A},\text{B} \right)
\end{equation}

Among competitors defined in Section~\ref{sec:def},
Adamic/Adar and EC have the same additivity property.
Katz does not have this property,
as in general (\ref{eq:katz}) in $G_3$ is more than the sum of that in $G_1$ and $G_2$,
and may even be divergent.

The additivity property is consistent with human intuition.
When multiple independent sets of evidence are combined,
our perception of proximity becomes the sum of proximity values derived from each individual set.
To state the same in more general terms, the proposed proximity measure (\ref{eq:metric}) is
proportional to the amount of evidence, which is why we choose it over (\ref{eq:nr}).
This additivity is also crucial to the development of
the approximation algorithm in Section~\ref{sec:algorithm}.

\emph{Monotonicity}.
Let $G_1 = \langle V_1,E_1 \rangle$ and $G_2 = \langle V_2,E_2 \rangle$ be two graphs
such that $V_1 \subseteq V_2$, $E_1 \subseteq E_2$, and that their weights satisfy
that $w_{V_1}\left( \textrm{X} \right) \leq w_{V_2}\left( \textrm{X} \right),\forall\textrm{X}\in V_1$
and that $w_{E_1}\left( e \right) \leq w_{E_2}\left( e \right),\forall e \in E_1$.
Let $s_{G_1} \left( \text{A},\text{B} \right)$ and $s_{G_2} \left( \text{A},\text{B} \right)$
be the measure value (\ref{eq:metric}) in these two graphs respectively.
Then the following condition holds.
\begin{equation} \label{eq:mono}
s_{G_1} \left( \text{A},\text{B} \right) \leq s_{G_2} \left( \text{A},\text{B} \right), \forall\textrm{A,B}\in V_1
\end{equation}

In plain language, if an edge is added to a graph or if a node/edge weight is increased in a graph,
then the proposed measure (\ref{eq:metric}) will not decrease for any pair of nodes.
This again is consistent with human intuition.

Among competitors defined in Section~\ref{sec:def},
Katz and EC have the same monotonicity property,
assuming that the additional edges or added weights do not cause (\ref{eq:katz}) to diverge.
In Adamic/Adar's (\ref{eq:adamicadar}), if the denominator is viewed as reciprocal of $w_V( \textrm{C} )$
which implies a specific choice\footnote{Such
choice of weights is shown to be beneficial in social networks \cite{adamicadar}.
With Blink Model, this can easily be encoded as domain knowledge, to be described in Section~\ref{sec:weight}.
In fact similar schemes are used in Section~\ref{sec:result}.} of $w_V$,
then it also satisfies the monotonicity property.
Note that (inverse) ERD is not monotonic because
additional edges may form a new long path from A to B and hence increase (\ref{eq:relidist}).

\subsection{Generalizations} \label{sec:gen}

The measure (\ref{eq:metric}) is defined on a particular event,
``a path exists from A to B''.
This definition is a pair-wise proximity measure and is
useful in for example link-prediction applications.
For other applications, the definition (\ref{eq:metric}) can be
generalized to other events in the blinking graph:
e.g., for a set of nodes $S_\textrm{A}$ and another set of nodes $S_\textrm{B}$,
\begin{equation} \label{eq:alt1}
\tilde{s} \left( S_\textrm{A},S_\textrm{B} \right) = - \log ( 1-\textrm{P} \left[ \textrm{ a path exists from any of } S_\textrm{A} \textrm{ to each of }S_\textrm{B} \right] )
\end{equation}
Or for three nodes A, B and C,
\begin{multline} \label{eq:alt3}
\tilde{\tilde{s}} \left( \textrm{A,B,C} \right) = - \log ( 1-\textrm{P} [ \textrm{ a path exists from A to B but no path exists}\\ \textrm{from A to C } ] )
\end{multline}
And there are many more possibilities.
In particular, when edges are labeled to indicate types of relations \cite{cbmm},
the choice of event can involve edge labels.

The measure (\ref{eq:metric}) is a proximity measure.
Another variation is a distance measure:
\begin{equation} \label{eq:dist}
d \left( \text{A},\text{B} \right) = - \log \left( b \left( \text{A},\text{B} \right) \right)
\end{equation}
It is straightforward to verify that the above definition satisfies the triangle inequality.
It also has the monotonicity property.
It has an additivity property that differs from that in Section~\ref{sec:prop},
but is defined on two graphs in series.

\subsection{Graph studies} \label{sec:graphs}

This section uses graph examples to compare the proposed proximity measure
with competitors to demonstrate that it is more consistent with human intuition.
Comparison with Adamic/Adar (\ref{eq:adamicadar}) or common-neighbor (replacing the sum
in (\ref{eq:adamicadar}) by a sum of 1's) is straightforward since they are limited to
length-2 paths, and an example is omitted.

\begin{figure}
\centering
\includegraphics[width=2.4in]{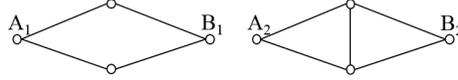}
\caption{A pair of graph examples.}
\label{fig:multi}
\end{figure}

\begin{figure}
\centering
\subfloat[]{\includegraphics[width=2.4in]{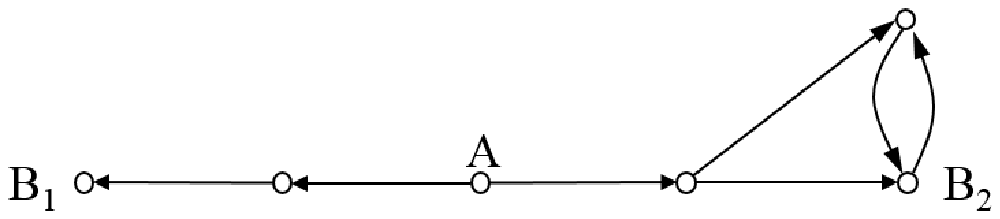}} \\
\subfloat[]{\includegraphics[width=2.4in]{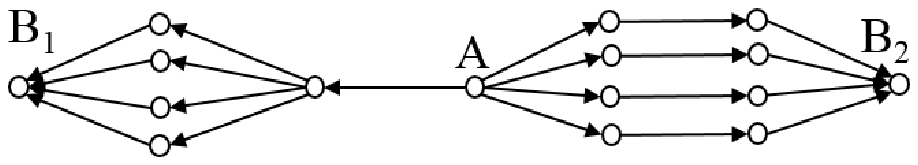}} \\
\subfloat[]{\includegraphics[width=2.1in]{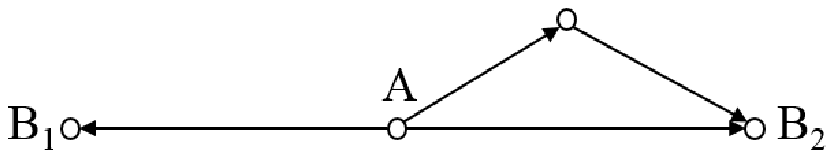}}
\caption{Graph examples for PPR, Katz and ERD.}
\label{fig:many}
\end{figure}

\begin{figure}
\centering
\includegraphics[width=3.3in]{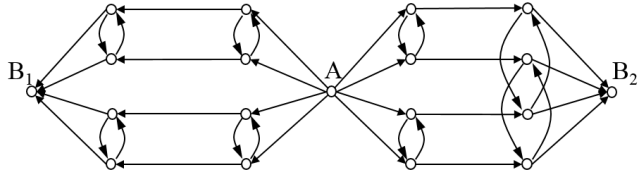}
\caption{A graph example.}
\label{fig:mark}
\end{figure}

In examples in Figures~\ref{fig:multi}--\ref{fig:mark},
we argue that human intuition would say that node A has stronger relation to node $\textrm{B}_2$ than to $\textrm{B}_1$.
A key notion is that human intuition not only prefers more and shorter paths,
but also prefers structures that are mutually corroborated.
If an edge or path has no corroboration, its existence in the graph may be a random
coincidence and hence does not indicate strong proximity.
On the flip side, proximity is strong for an aggregate structure that is impervious to edges randomly existing.

In discussing all examples, we assume uniform node weights of 1 and uniform edge weights of $w<1$.

Let us begin with Figure~\ref{fig:multi} of two undirected graphs.
It could be perceived that there are two random paths from $\textrm{A}_1$ to $\textrm{B}_1$,
while the two length-2 paths from $\textrm{A}_2$ to $\textrm{B}_2$ are less likely to be random because
the crossing edge provides mutual corroboration between them,
and therefore human intuition prefers $(\textrm{A}_2,\textrm{B}_2)$ over $(\textrm{A}_1,\textrm{B}_1)$.
Table~\ref{tbl:multi} lists various proximity scores, where none is consistent with human intuition.
Shortest-path and EC conclude that $(\textrm{A}_1,\textrm{B}_1)$ and $(\textrm{A}_2,\textrm{B}_2)$ are equally related,
while CFEC \cite{koren2006} and commute-time \cite{fouss} conclude that $(\textrm{A}_1,\textrm{B}_1)$ is stronger than $(\textrm{A}_2,\textrm{B}_2)$.
In contrast, the Blink Model score is $-2\cdot\log(1-w^2)$ for $(\textrm{A}_1,\textrm{B}_1)$
and $-\log(1-2w^2-2w^3+5w^4-2w^5)$ for $(\textrm{A}_2,\textrm{B}_2)$, and the latter is strictly larger than the former.
This shows a weakness of EC in that it sees no effect from the crossing edge in the second graph;
the EC variant of CFEC \cite{koren2006} exacerbates this trait;
another EC variant \cite{faloutsos,tong} adds a universal sink to the EC model,
and it is straightforward to verify that it also ranks $(\textrm{A}_1,\textrm{B}_1)$ as stronger than $(\textrm{A}_2,\textrm{B}_2)$,
and similar effects of the universal sink have been reported in \cite{koren2006}.
A spectrum of measures was proposed in \cite{yen}, where shortest-path is one end of the spectrum
while commute-time is the other end;
although we are unable to judge intermediate variants of \cite{yen},
Table~\ref{tbl:multi} suggests that both of its corner variants produce counterintuitive rankings for Figure~\ref{fig:multi}.

\begin{table}[t]
\centering
\caption{Some proximity measures on Figure~\ref{fig:multi}.}
\label{tbl:multi}
\small
\begin{tabular}{|l|c|c|} \hline
Measure         & $\textrm{A}_1$,$\textrm{B}_1$ & $\textrm{A}_2$,$\textrm{B}_2$ \\ \hline
1/shortest-path & $1/(2w)$         & $1/(2w)$         \\ \hline
1/commute-time  & $1/(8w)$         & $1/(10w)$        \\ \hline
EC              & $w$              & $w$              \\ \hline
CFEC            & $w$              & $8w/9$           \\ \hline
\end{tabular}
\end{table}

Next let us consider Figure~\ref{fig:many}(i).
There are two equal-length paths from A to $\textrm{B}_1$ and to $\textrm{B}_2$,
but there are more paths going from A to $\textrm{B}_2$.
So it seems there's more reason to believe that it's not a coincidence
that $\textrm{B}_2$ is connected to A than $\textrm{B}_1$.
PageRank scores are $\text{score}_\text{PPR} \left( \textrm{A,B}_1 \right) = (1-\alpha)^2/2$
and $\text{score}_\text{PPR} \left( \textrm{A,B}_2 \right) = (1-\alpha)^2/4$.
In other words, PPR considers that A has greater proximity to $\textrm{B}_1$ than to $\textrm{B}_2$,
and this holds true for any parameterization.
In contrast, the Blink Model score (\ref{eq:metric}) is higher for $\textrm{B}_2$ than $\textrm{B}_1$.

Consider the Katz measure (\ref{eq:katz}) on Figure~\ref{fig:many}(ii).
It's straightforward to verify that, with any $w$ and $\beta$ values,
we have $\text{score}_\text{Katz} \left( \textrm{A,B}_1 \right)=\text{score}_\text{Katz}\left( \textrm{A,B}_2 \right)$,
including when $w$ is 1 and (\ref{eq:katz}) becomes the original Katz.
In other words, the (modified) Katz measure cannot discern $\textrm{B}_1$ and $\textrm{B}_2$ relative to A,
because it sees no difference between the four paths to $\textrm{B}_1$ and those to $\textrm{B}_2$.
In contrast, the Blink Model measure (\ref{eq:metric}) recognizes that
the edge to the left of A, which all paths to $\textrm{B}_1$ depend on, has no corroboration,
and we have $s \left( \text{A},\text{B}_1 \right) < s \left( \text{A},\text{B}_2 \right)$,
consistent with human intuition.

Next consider the ERD measure (\ref{eq:relidist}) on Figure~\ref{fig:many}(iii),
and we have $\text{score}_\text{ERD} \left( \textrm{A,B}_1 \right) = 1$
and $\text{score}_\text{ERD} \left( \textrm{A,B}_2 \right) > 1$.
Since ERD is an inverse proximity, the conclusion is that 
A has greater proximity to $\textrm{B}_1$ than to $\textrm{B}_2$,
and is inconsistent with human intuition.
Blink Model shows the reverse.

Next consider SimRank \cite{simrank} on a three-node undirected complete graph
and on a four-node undirected complete graph.
It is straightforward to verify that the SimRank score, under any parameterization,
is higher for a pair of nodes in the former than in the latter,
and in fact the score always decreases as the size of a complete graph increases. 
This is contrary to our Blink Model and human intuition.
To be fair, SimRank was designed as a similarity measure and was not intended
to be a proximity measure. This is also the likely reason that
its performance in \cite{liben} was reported to be mediocre.

The last example graph is Figure~\ref{fig:mark}, which demonstrates
the advantage of measure (\ref{eq:metric}) over a class of methods.
Continuing the intuition from Figure~\ref{fig:multi},
on the left there exists mutual corroboration between the top pair of length-3 paths to $\text{B}_1$
and between the bottom pair of length-3 paths, but none exists between the two pairs.
On the right there exists mutual corroboration among all four length-3 paths to $\text{B}_2$,
and the proximity to $\text{B}_2$ is perceived as more robust to a human.
This is analogous to using four strings to reinforce four poles, and a human
is more likely to use a pattern similar to the right half of Figure~\ref{fig:mark} than the left.
The Blink Model recognizes that $\text{B}_2$ is more connected to A than $\text{B}_1$,
e.g. when $w$ is 0.5, $s \left( \text{A},\text{B}_1 \right) = 0.795$
and $s \left( \text{A},\text{B}_2 \right) = 0.809$.

Consider any algorithm which operates on local storage per node:
it starts with special storage in source node A, all other nodes starting equal;
on each successive iteration, it updates information stored in each node
based on information stored in adjacent nodes from the previous iteration.
Such an algorithm can easily for example compute shortest-path distance from A,
the PPR score, the Katz score, and the EC score.
However, such an algorithm, even with an infinite amount of storage and an infinite number of iterations,
cannot distinguish $\text{B}_1$ and $\text{B}_2$ in Figure~\ref{fig:mark};
in fact, the eight nodes that are distance-2 from A are also indistinguishable.
Algorithms of this type encompass a large range of methods.
In particular, any matrix computation based on the adjacency matrix or variants of the adjacency matrix,
which includes almost all measures that have a random-walk-based definition,
falls into this category,
and no possible linear algebra can determine that $\text{B}_2$ is closer than $\text{B}_1$.
This is a blessing and a curse: the Blink Model can discern cases correctly,
but is inherently hard to compute.

\subsection{Training weights} \label{sec:weight}

This section addresses a practical issue of using the Blink Model in an application:
how to set edge and node weights.
There are many ways, and we describe a simple yet practical method to do so by training a few parameters.

Let two functions $f_E:E \rightarrow R_{>0}$ and $f_V:V \rightarrow R_{>0}$ represent domain knowledge.
In applications where we have no domain knowledge beyond the topology $G$, we simply have $f_E$ and $f_V$ equal to 1 for all.
In applications where we do, we assume that $f_E$ and $f_V$ are larger for more important or more reliable edges and nodes,
and that their values exhibit linearity:
two parallel edges $e_1$ and $e_2$ can be replaced by a single edge $e_3$ with $f_E(e_3)=f_E(e_1)+f_E(e_2)$.

Our method sets graph edge and node weights as:
\begin{equation}  \label{eq:weight}
\begin{aligned}
w_E(e) & = 1 - \left( 1 - b_1 \right) ^ {f_E(e)}, \forall e \in E \\
w_V(v) & = 1 - \left( 1 - b_2 \right) ^ {f_V(v)}, \forall v \in V
\end{aligned}
\end{equation}
where $b_1,b_2\in(0,1)$ are two tunable parameters\footnote{One way to interpret $b_1$ and $b_2$,
or $w_E$ and $w_V$ in general, is that they encode varying personalities, while
the analysis engine (\ref{eq:metric}) is invariant. This is similar to that different
people, when presented with the same evidence, may make different decisions.
In Section~\ref{sec:result} for example, we scan for $b_1$ and $b_2$ that match the collective
behavior of physicists or Wikipedia contributors.}.
It is straightforward to verify that the linearity assumption on $f_E$
is consistent with the arithmetic in Section~\ref{sec:basic}.
Parameters $b_1$ and $b_2$ for Blink Model are similar to $\alpha$ for PageRank and $\beta$ for Katz,
and we search for best values by using training data in an application.
If $f_E$ and $f_V$ have tunable parameters, those can be trained in the same process.
Since we introduce only two parameters, the training process is straightforward
and can be brute-force scan and/or greedy search.

This method is applicable to all proximity measures and is used for all in Section~\ref{sec:result}.
One caveat is that certain measures work better with linear weights:
\begin{equation}  \label{eq:linearweight}
\begin{aligned}
w_E(e) & = b_1 \cdot f_E(e), \forall e \in E \\
w_V(v) & = b_2 \cdot f_V(v), \forall v \in V
\end{aligned}
\end{equation}
For example, we observe empirically that PPR works better with (\ref{eq:linearweight}),
which is intuitive given the linearity assumption on $f_E$,
while Modified Katz and ERD prefer (\ref{eq:weight}).
Note that when $b_1$ and $b_2$ are small, (\ref{eq:weight}) asymptotically becomes (\ref{eq:linearweight}).

\section{Approximation Algorithm} \label{sec:algorithm}

We present a deterministic algorithm
that approximates (\ref{eq:metric}) directly.
Without loss of generality (per Section~\ref{sec:basic}),
we describe this algorithm under the conditions of all node weights being 1
and that $G$ is a simple directed graph where parallel edges are already merged.

\subsection{Overall flow} \label{sec:flow}

A \emph{minimal path} from node A to node B is defined as a path from A to B without repeating nodes.
For a finite graph $G$, there exist a finite number of minimal paths.
Consider a single minimal path $i$ from node A to node B. We define the following\footnote{We
omit A and B from $s_{\textrm{path }i}$ and $\hat{s}_{\textrm{path }i}$ notations to keep formulas concise.
Note that any path refers to a minimal path in $G$ from A to B.}
as the \emph{nominal contribution} of this path to (\ref{eq:metric}).
\begin{equation} \label{eq:nominal}
s_{\textrm{path }i} = - \log \left( 1 - \prod_{\textrm{edge }e\textrm{ on path }i}w_E\left(e\right) \right)
\end{equation}
By the additivity property (\ref{eq:add}), if all minimal paths from A to B are mutually disjoint,
we can compute (\ref{eq:metric}) exactly by summing (\ref{eq:nominal}) over all minimal paths.
Of course this is not true for general graphs where paths from A to B overlap each other
and (\ref{eq:metric}) is less than the sum of $s_{\textrm{path }i}$ values.

However, if we consider only length-1 and length-2 minimal paths, they can never
share an edge with each other, and their nominal contributions can be added according to the additivity property.
Further invoking the monotonicity property (\ref{eq:mono}), we obtain the following inequality.
\begin{equation} \label{eq:bounds}
\sum_{\textrm{path }i\textrm{ with length 1 or 2}} {s_{\textrm{path }i}}
\leq s \left( \text{A},\text{B} \right)
\leq \sum_{\textrm{path }i} {s_{\textrm{path }i}}
\end{equation}
Therefore, the key to approximate (\ref{eq:metric}) is to quantify the contribution
of minimal paths that are longer than 2.

We start the approximation by making the following conjecture that the contribution
of each minimal path $i$ is quantifiable as a value $\hat{s}_{\textrm{path }i}$ such that
$s \left( \text{A},\text{B} \right) = \sum_{\textrm{path }i} {\hat{s}_{\textrm{path }i}}$.
We use $G^\prime$ to denote a subgraph of $G$
and $s_{G^\prime} \left( \text{A},\text{B} \right)$ to denote the measure value (\ref{eq:metric}) in $G^\prime$.

\newtheorem{assumption}{Conjecture}
\begin{assumption}\label{assume}
A value $\hat{s}_{\textrm{path }i}$ exists for each minimal path $i$ from node A to node B,
such that these $\hat{s}_{\textrm{path }i}$ values satisfy the following conditions:
\begin{equation} \label{eq:assumption}
\begin{aligned}
\hat{s}_{\textrm{path }i} = s_{\textrm{path }i}, & \textrm{ if path }i\textrm{ has length 1 or 2} \\
0 \leq \hat{s}_{\textrm{path }i} \leq s_{\textrm{path }i}, & \textrm{ if path }i\textrm{ is longer than 2}
\end{aligned}
\end{equation}
\begin{equation} \label{eq:assumption2}
\begin{aligned}
s_{G^\prime} \left( \text{A},\text{B} \right) & \geq \sum_{\textrm{path }i\textrm{ is contained in }G^\prime} {\hat{s}_{\textrm{path }i}} \\
s_{G^\prime} \left( \text{A},\text{B} \right) & \leq \sum_{\textrm{path }i\textrm{ overlaps with }G^\prime} {\hat{s}_{\textrm{path }i}}
\end{aligned}
\end{equation}
for any subgraph $G^\prime$.
\end{assumption}

Our algorithm works best when Conjecture~\ref{assume} holds,
while the approximation would be coarser when it does not.
We have not found any graph that breaks Conjecture~\ref{assume},
and it remains unproven whether it holds for all graphs.
We use Conjecture~\ref{assume} in two ways.
By selecting a special set of subgraphs $G^\prime$, we utilize (\ref{eq:assumption2})
to iteratively approximate $\hat{s}_{\textrm{path }i}$ values.
Then, after obtaining approximate $\hat{s}_{\textrm{path }i}$ values,
we invoke Conjecture~\ref{assume} for a special case of $G^\prime=G$,
where the two sums in condition (\ref{eq:assumption2}) are identical
and therefore (\ref{eq:assumption2}) becomes two equalities.
This justifies that $s \left( \text{A},\text{B} \right) = \sum_{\textrm{path }i} {\hat{s}_{\textrm{path }i}}$
which achieves our purpose.

One observation is that Conjecture~\ref{assume} does not uniquely define $\hat{s}_{\textrm{path }i}$ values
as there may exist two sets of $\hat{s}_{\textrm{path }i}$ values that both
satisfy (\ref{eq:assumption})(\ref{eq:assumption2}).
However, by definition they both sum up to the same end result (\ref{eq:metric}),
and therefore we only need to find one such set of $\hat{s}_{\textrm{path }i}$ values.
A second observation is that the lower bound in (\ref{eq:assumption2}) is tight for a variety of subgraphs $G^\prime$,
while the upper bound is tight only for large subgraphs.
We exploit this observation in the proposed algorithm:
we will select/design a certain set of subgraphs $G^\prime$ where the lower bound in (\ref{eq:assumption2}) is tight,
and then use the lower bound as an equality to iteratively refine the approximated $\hat{s}_{\textrm{path }i}$ values.

The proposed algorithm is illustrated in Algorithm~\ref{pseudo}.
$\hat{s}_{\textrm{path }i}$ values are initialized to be the nominal.
A subgraph $G^\prime_i$, to be elaborated later, is selected for each path $i$ longer than 2,
and $s_{G^\prime_i} \left( \text{A},\text{B} \right)$ is computed/approximated.
During each iteration, for each path $i$ longer than 2,
we use the lower bound in (\ref{eq:assumption2}) as an equality and convert it to the following update formula.
\begin{equation} \label{eq:update}
\hat{s}_{\textrm{path }i}^{k+1} = \frac{\hat{s}_{\textrm{path }i}^k  \cdot \left( s_{G^\prime_i} \left( \text{A},\text{B} \right) - \sum_{\textrm{path }j\in\Upsilon_i} {s_{\textrm{path }j}} \right)}{\sum_{\textrm{path }j\in\Xi_i} {\hat{s}_{\textrm{path }j}^k}}
\end{equation}
where $k$ is the iteration index,
$\Xi_i$ is the set of minimal paths from A to B that are contained in $G^\prime_i$
and that have length more than 2,
and $\Upsilon_i$ is the set of minimal paths from A to B that are contained in $G^\prime_i$
and that have length of 1 or 2.

\newtheorem{algorithm}{Algorithm}
\begin{algorithm}
Overall flow of the proposed algorithm.
\begin{center}
\begin{tabular}{l}
\hline
\texttt{\upshape $\hat{s}^0_{\textrm{path }i} \gets s_{\textrm{path }i}$, $\forall$ path $i$, using (\ref{eq:nominal}); }
\\
\texttt{\upshape select subgraphs $G^\prime_i$, $\forall$ path $i$ longer than 2;}
\\
\texttt{\upshape for $k=0,1,\cdots$, until convergence:}
\\
$\quad$ \texttt{\upshape compute $\hat{s}_{\textrm{path }i}^{k+1}$ using (\ref{eq:update}), $\forall$ path $i$ longer than 2;}
\\
\texttt{\upshape $s \left( \text{A},\text{B} \right) \gets$ sum of final $\hat{s}_{\textrm{path }i}$ values;}
\\
\hline
\end{tabular}
\end{center}
\label{pseudo}
\end{algorithm}

One way to interpret this algorithm is that it is an iterative solver that solves a linear system
where there is one equation for each $G^\prime_i$ and the unknowns are $\hat{s}_{\textrm{path }i}$.
This interpretation holds for the variation in Section~\ref{sec:high},
however it does not hold in Sections~\ref{sec:mid} and \ref{sec:low},
in which we will present an alternative interpretation.
In the next three sections, we present three variations of the proposed algorithm.
They differ in how they select/construct subgraph $G^\prime_i$ for a given path $i$.
Section~\ref{sec:paths} discusses selecting minimal paths to work on.

\begin{figure}
\centering
\subfloat[]{\includegraphics[width=1.6in]{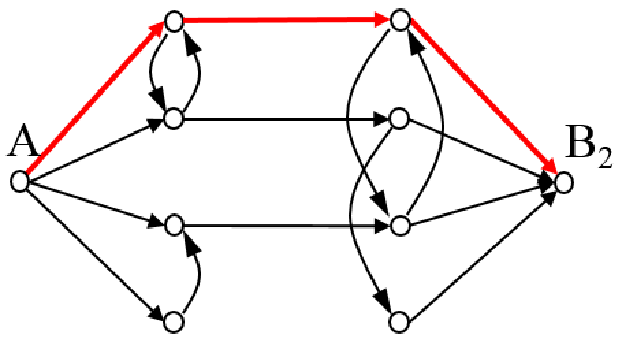}}
\subfloat[]{\includegraphics[width=1.6in]{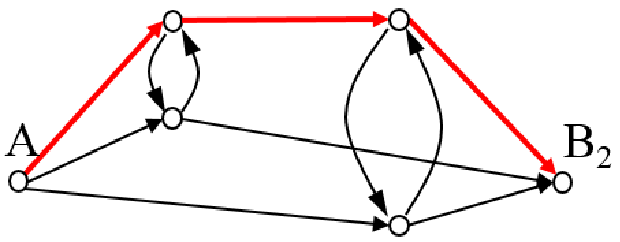}}
\caption{(i) The selected subgraph of Figure~\ref{fig:mark} 
for the highlighted path, and (ii) its simplified form.}
\label{fig:subgraph}
\end{figure}

\subsection{High-accuracy variation} \label{sec:high}

In this variation of the proposed algorithm,
we select subgraph $G^\prime_i$ as the minimal subgraph
that contains all minimal paths from A to B which overlap with path $i$ by at least one edge.
One example is illustrated in Figure~\ref{fig:subgraph}(i),
which shows the subgraph of Figure~\ref{fig:mark} 
for the highlighted path from A to $\textrm{B}_2$,
and it is used in (\ref{eq:update})  to update
$\hat{s}$ of the highlighted path during each iteration.
Note that $G^\prime_i$ only needs to be identified once
and $s_{G^\prime_i} \left( \text{A},\text{B} \right)$
only needs to be evaluated once, and the same value
is reused in (\ref{eq:update}) across iterations.

A main computation in this variation is the evaluation of
$s_{G^\prime_i} \left( \text{A},\text{B} \right)$.
Since $G^\prime_i$ is much smaller than the whole graph $G$ in
typical applications, many techniques from the network reliability field
can be applied to approximately evaluate (\ref{eq:nr}) in $G^\prime_i$
and hence $s_{G^\prime_i} \left( \text{A},\text{B} \right)$.
For example, it is known that (\ref{eq:nr}) is invariant under
certain topological transformations \cite{ball,satyanarayana}.
Applying such transformations, the graph in Figure~\ref{fig:subgraph}(i)
can be simplified to Figure~\ref{fig:subgraph}(ii) without loss of accuracy.
Then a Monte Carlo method can be applied on the simplified graph
and approximate $s_{G^\prime_i} \left( \text{A},\text{B} \right)$.

\subsection{Medium-accuracy variation} \label{sec:mid}

Instead of identifying and solving each $G^\prime_i$ as an actual subgraph,
in this variation we construct $G^\prime_i$ as a hypothetical subgraph for each path $i$
during each iteration.

We start the construction by considering the
amount of sharing on each edge.
In the $k^\textrm{th}$ iteration, for edge $e$, define
\begin{equation} \label{eq:usage}
u_e^k = \sum_{\textrm{path }j\textrm{ contains }e\textrm{ and is longer than 2}}{\hat{s}_{\textrm{path }j}^k}
\end{equation}
Intuitively, $u_e^k$ quantifies usage of edge $e$ by A-to-B minimal paths,
based on current knowledge at the $k^\textrm{th}$ iteration.

\begin{figure}
\centering
\subfloat[]{\includegraphics[width=2.6in]{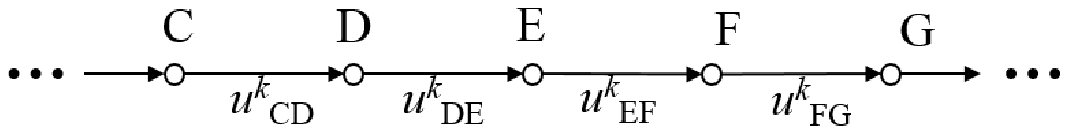}}\\
\subfloat[]{\includegraphics[width=2.6in]{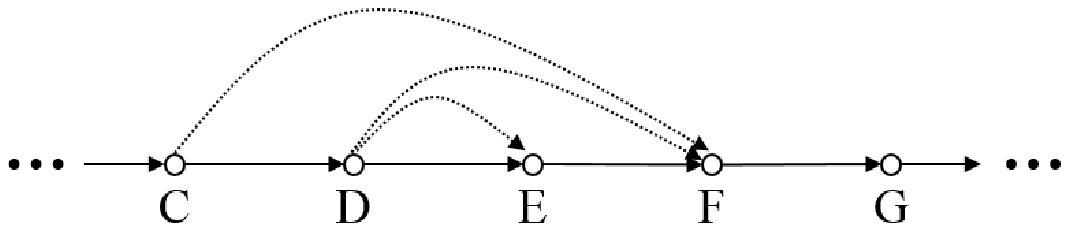}}
\caption{(i) Example of a middle section of a path.
(ii) The same middle section after adding hypothetical edges.}
\label{fig:path}
\end{figure}

For each path $i$, we annotate each edge on this path with $u_e^k$.
Figure~\ref{fig:path}(i) illustrates a middle section of an example path $i$.
We use $u_{\textrm{XY}}^k$ to denote $u_e^k$ when $e$ is an edge from node X to node Y,
and $w_{\textrm{XY}}$ to denote its edge weight.
Without loss of generality, we assume that
$u_{\textrm{FG}}^k > u_{\textrm{CD}}^k > u_{\textrm{EF}}^k > u_{\textrm{DE}}^k$.

We construct the hypothetical subgraph $G^\prime_i$ starting from path $i$ itself
and by adding hypothetical edges.
Since $u_{\textrm{EF}}^k > u_{\textrm{DE}}^k$, there must exist one or more A-to-B path(s)
that passes the edge from E to F but that does not
pass the edge from D to E. A hypothetical new edge from D to E is added
to approximate the effect of such path(s);
furthermore, we know that the sum of $\hat{s}^k$ of these paths is
equal to $u_{\textrm{EF}}^k - u_{\textrm{DE}}^k$, and we use this fact to assign
the following weight.
\begin{equation} \label{eq:hypoweight1}
w_\textrm{DE}^\prime = 1 - \left( 1 - w_\textrm{DE} \right)^{\left(u_{\textrm{EF}}^k - u_{\textrm{DE}}^k\right)/ u_{\textrm{DE}}^k}
\end{equation}
In plain words, we assume that this hypothetical edge
is $\left( u_{\textrm{EF}}^k - u_{\textrm{DE}}^k \right)/u_{\textrm{DE}}^k$
times as strong as original D-to-E edge.

Similarly, since $u_{\textrm{CD}}^k > u_{\textrm{EF}}^k$, there must exist one or
more A-to-B path(s) that passes the edge from C to D, but that does not pass the edge
from E to F, the edge from D to E, or paths represented by the hypothetical edge from
D to E. A hypothetical new edge from D to F is added to approximate the
effect of such path(s). Again, we know that the sum of $\hat{s}^k$ of these paths is
equal to $u_{\textrm{CD}}^k - u_{\textrm{EF}}^k$, and by the same argument for (\ref{eq:hypoweight1}),
we assign the following edge weight.
\begin{equation} \label{eq:hypoweight2}
w_\textrm{DF}^\prime = 1- \left( 1 - w_\textrm{EF} \cdot \left( 1 - \left( 1 - w_\textrm{DE} \right)^{u_{\textrm{EF}}^k / u_{\textrm{DE}}^k} \right) \right)^\frac{u_{\textrm{CD}}^k - u_{\textrm{EF}}^k}{u_{\textrm{EF}}^k}
\end{equation}
The same rationale applies to adding the hypothetical edge from C to F, and so on.

The above construction process for $G^\prime_i$ processes edges on path $i$ one by one
in the order of increasing $u_e^k$ 
values and adds hypothetical edges.
The last step of construction is to add to $G^\prime_i$
the length-2 A-to-B paths that overlap with path $i$,
since they are not visible in $u_e^k$
values.
The completed hypothetical subgraph $G^\prime_i$ is then used
in (\ref{eq:update})  to compute $\hat{s}_{\textrm{path }i}^{k+1}$ for the next iteration,
and the overall algorithm proceeds.
Note that the denominator $\sum_{\textrm{path }j\in\Xi_i} {\hat{s}_{\textrm{path }j}^k}$ in (\ref{eq:update})
is simply equal to the largest $u_e^k$ along path $i$; let it be $u_\textrm{max}^k$.

One distinction between this variation and Section~\ref{sec:high}
is that the exact evaluation of $s_{G^\prime_i} \left( \text{A},\text{B} \right)$
has linear complexity with respect to path length.
The hypothetical edges form series-parallel structures that
are friendly to topological transformations \cite{satyanarayana,politof}.
Using Figure~\ref{fig:path}(ii) as an example,
the hypothetical D-to-E edge and the original D-to-E edge can be
merged into a single edge;
then it and the E-to-F edge can be merged into a single D-to-F edge;
then it and the hypothetical D-to-F edge can be merged, and so on.

Another distinction between this variation and Section~\ref{sec:high}
is that $G^\prime_i$ is no longer the same across iterations.
As a result, the linear-system interpretation mentioned in Section~\ref{sec:flow}
 no longer holds.
Instead, the following interpretation is more intuitive.
The calculation by (\ref{eq:update}) 
applies a dilution factor $\hat{s}_{\textrm{path }i}^k/u_\textrm{max}^k$
on the strength of $G^\prime_i$ excluding length-2 paths.
The more path $i$ overlaps with other paths,
the larger $u_\textrm{max}^k$ is and the smaller the dilution factor is,
and $G^\prime_i$ is a hypothetical subgraph that mimics a path where every edge has usage $u_\textrm{max}^k$.

\begin{figure}
\centering
\includegraphics[width=1.9in]{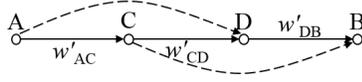}
\caption{Hypothetical subgraph in low-accuracy variation.}
\label{fig:low}
\end{figure}

\subsection{Low-accuracy variation} \label{sec:low}

Continuing the interpretation from the last section,
we may construct $G^\prime_i$ in the form
of Figure~\ref{fig:low}, which results in a tradeoff
with further lower accuracy and lower runtime.

For minimal path $i$, let C and D be the first and last intermediate nodes,
let the original edge weights along this path be $w_1$, $w_2$, $\cdots$, $w_n$,
and let edge usages (\ref{eq:usage})
along this path be $u_1^k$, $u_2^k$, $\cdots$, $u_n^k$.
We construct a hypothetical subgraph $G^\prime_i$ in the form of Figure~\ref{fig:low}.
In Figure~\ref{fig:low}, dashed arrows from A to D and from C to B represent edges
that may exist in $G$ and form length-2 paths A-D-B and A-C-B;
if these dashed edges do exist, they have their original weights.
The weights of the three solid edges are:
\begin{eqnarray} \label{eq:low}
w_\textrm{AC}^\prime & = & 1- \left( 1 - w_1 \right)^{u_\textrm{max}^k / u_1^k} \\
w_\textrm{CD}^\prime & = & 1- \left( 1 - \prod_{j=2}^{n-1}{w_j} \right)^{u_\textrm{max}^k / u_\textrm{mean}^k} \\
w_\textrm{DB}^\prime & = & 1- \left( 1 - w_n \right)^{u_\textrm{max}^k / u_n^k}
\end{eqnarray}
where $u_\textrm{mean}^k$ is a weighted average of $u_2^k,\cdots,u_{n-1}^k$:
\begin{equation} \label{eq:mean}
u_\textrm{mean}^k = \left( \sum_{j=2}^{n-1}{u_j^k\cdot\log\left(w_j\right)} \right) / {\sum_{j=2}^{n-1}{\log\left(w_j\right)}}
\end{equation}
Intuitively, $G^\prime_i$ still mimics a path where every edge has usage $u_\textrm{max}^k$,
and is constructed more crudely than in Section~\ref{sec:mid}.

\begin{table*}[!t]
\centering
\caption{Performance on Figure~\ref{fig:mark}.
$\textrm{Error}_{\text{B}_2}$ and $\textrm{Error}_\Delta$ are relative errors in $s \left( \text{A},\text{B}_2 \right)$
and in $s \left( \text{A},\text{B}_2 \right) - s \left( \text{A},\text{B}_1 \right)$.}
\label{tbl:toy}
\footnotesize
\setlength\tabcolsep{1pt}
\begin{tabular}{|l|c|c|c|c|c|c|c|c|c|} \hline
Edge weight & \multicolumn{3}{|c|}{0.1} & \multicolumn{3}{|c|}{0.5} & \multicolumn{3}{|c|}{0.9} \\ \hline
& $\textrm{Error}_{\text{B}_2}$ & $\textrm{Error}_\Delta$ & Runtime(s)
& $\textrm{Error}_{\text{B}_2}$ & $\textrm{Error}_\Delta$ & Runtime(s)
& $\textrm{Error}_{\text{B}_2}$ & $\textrm{Error}_\Delta$ & Runtime(s) \\ \hline
High accuracy   & 0.07\% & 19.2\% & 7.68E-3 & 2.40\% & 12.4\% & 6.73E-3 & 2.40\% & 6.73\% & 7.74E-3 \\ \hline
Medium accuracy & 8.86\% & 100\%  & 1.48E-3 & 17.2\% & 100\%  & 1.50E-3 & 12.0\% & 100\%  & 4.78E-4 \\ \hline
MC, 1K samples        & 37.7\% & 4.21E2 & 9.53E-4 & 3.38\% & 282\%  & 1.94E-3 & 43.1\% & 228\%  & 2.09E-3 \\ \hline
MC, 10K samples       & 12.9\% & 1.56E2 & 8.45E-3 & 1.07\% & 106\%  & 1.84E-2 & 3.82\% & 272\%  & 1.91E-2 \\ \hline
MC, 100K samples      & 3.55\% & 4751\% & 8.27E-2 & 0.35\% & 29.2\% & 0.174   & 1.05\% & 70.2\% & 0.182   \\ \hline
MC, 1M samples        & 1.13\% & 1852\% & 0.812   & 0.11\% & 8.69\% & 1.74    & 0.33\% & 23.7\% & 1.78    \\ \hline
\end{tabular}
\end{table*}

\subsection{Accuracy and implementation issues} \label{sec:paths}

This algorithm can discern $\text{B}_1$ and $\text{B}_2$ correctly for all cases in Section~\ref{sec:graphs},
and Table~\ref{tbl:toy} shows details on Figure~\ref{fig:mark},
with three cases where all edge weights are 0.1, 0.5 and 0.9 respectively.
The medium-accuracy variation is unable to distinguish $\text{B}_1$ and $\text{B}_2$
and hence has 100\% $\textrm{Error}_\Delta$.
Each Monte Carlo (MC) measurement is repeated with 100 different random number
generation seeds, and the reported error/runtime is the average over the 100 runs.
Not surprisingly, MC favors 0.5 edge weight and has larger errors for higher or lower weights,
while our algorithm is stable across the range.
Table~\ref{tbl:toy} suggests that the high-accuracy variation
has comparable accuracy to 10K MC samples for individual-score estimation,
and is more accurate in differentiating $\text{B}_1$ and $\text{B}_2$ than one million samples for two of the three cases.
It also suggests that a MC method needs at least tens of thousands of samples to reliably differentiate nodes.

For this graph, the high-accuracy runtime is equal to 3500--8000 MC samples,
while the medium-accuracy runtime is equal to 200--1500 samples.
In practice, different variations can form a hybrid: medium- or even low-accuracy variation is used
for all target nodes and identifies a critical subset of targets,
while high-accuracy variation only ranks within the critical subset.

With any variation, the input to the proposed algorithm is a set of minimal paths from A to B,
and the output are $\hat{s}$ values for each path in the set.
For large graphs, to maintain practical complexity, the set is a subset of all minimal paths,
and this results in underestimation of the Blink Model score (\ref{eq:metric}).
A natural strategy is to ignore long and weak paths, similar to \cite{liu} with respect to PageRank.
This is motivated by the fact that the measure (\ref{eq:nr}) has locality \cite{karger}.
In our implementation, a minimal path $i$ is given to the proposed algorithm
if and only if it satisfies both of the following conditions:
\begin{equation} \label{eq:filter1}
s_{\textrm{path }i} \ge t_1
\end{equation}
\begin{equation} \label{eq:filter2}
\prod_{\textrm{edge }e\textrm{ on path }i}{\frac{\log\left(1-w_E\left(e\right)\right)}{\sum_{\textrm{edge }f\in\Theta_e}{\log\left(1-w_E\left(f\right)\right)}}} \ge t_2
\end{equation}
where $\Theta_e$ is the set of out-going edges from the source node of $e$,
and $t_1$ and $t_2$ are two constant thresholds that control runtime-accuracy tradeoff.
Condition~(\ref{eq:filter2}) is essentially a fan-out limit on paths.
When making predictions in Section~\ref{sec:result},
we use $t_2$$=$2E-6 which implies that we consider
at most 500,000 paths from A.\footnote{Identification
of minimal paths to many B's can be achieved by
a single graph traversal from A. For example, if the traversal finds a path
composed of nodes A, $\textrm{B}_1$, $\textrm{B}_2$, $\cdots$, $\textrm{B}_n$,
which satisfy (\ref{eq:filter1})(\ref{eq:filter2}), then this provides one qualified path
to $\textrm{B}_1$, one to $\textrm{B}_2$, $\cdots$, and one to $\textrm{B}_n$.}
For node B that is close to A, the above strategy provides good coverage of
a ``local'' region and thus causes little underestimation.
The further away B is from A, the less complete this coverage is and
the more the underestimation is.
In applications where we rank multiple B nodes for a given A, e.g. link prediction,
fidelity is maintained because the degree of underestimation
is negatively correlated with exact scores.
For a distant node B where no path satisfies the bounds,
we use a single path with the largest $s_{\textrm{path }i}$, which can be found
efficiently with a Dijkstra's-like procedure.

On a last note, our algorithm is essentially a method to quantify values
for a set of uncertain and mutually overlapping pieces of evidence, each piece being a minimal path.
The algorithm is orthogonal to the choice of input evidence,
for which any path collection implementation can be used,
while the quality of output is influenced by the quality of input.

\section{Experimental Results} \label{sec:result}

This section compares the predictive power of our method
against competitors on two temporal link prediction tasks.
Data and benchmarking code are released at \cite{data}.
All blink-model runs use the variation of Section~\ref{sec:mid};
single-thread run time is 2.9--5.0 seconds per starting node in coauthorship graphs
and 5.3 seconds in the Wikipedia graph,
on a Linux server with Intel E7-8837 processors at 2.67GHz.

We use the method of Section~\ref{sec:weight} with two scenarios:
graph \#1 is with no domain knowledge, $f_E$ and $f_V$ being 1 for all,
and hence with uniform edge and node weights $b_1$ and $b_2$;
graph \#2 is with domain knowledge expressed in $f_E$ and $f_V$.
In Section~\ref{sec:arxiv}, we follow the practice of \cite{liben} and scan parameters for each
predictor without separating training and test sets.
In Section~\ref{sec:wikipedia}, we separate data into a training set and a test set,
and the training set is used to scan for the best parameterization,
while the test set is used for evaluation.
Best parameter values for each predictor are listed in Tables \ref{tbl:arxiv} and \ref{tbl:wiki};
no-effect parameters are omitted, e.g., PPR scores are invariant with respect to $b_1$ and $b_2$.

We focus on evaluating individual proximity measures and do not compare with ensemble methods \cite{lichtenwalter}.
We use structured input data as is, and do not utilize text of arXiv papers or Wikipedia pages.
In real life applications, natural language processing (NLP) 
can be used to provide additional edges,
edge labels and more meaningful edge weights \cite{cbmm}, and thereby enhance prediction accuracy.
Furthermore, such data from NLP are inherently noisy,
which fits perfectly with the underlying philosophy of the Blink Model that any evidence is uncertain.

\begin{table*}[ht]
\centering
\caption{Statistics of the coauthorship networks. Entry
format is our-number/number-reported-in-\cite{liben}.
Column Collaborations denotes pairwise relations in the training period.
Column $|E_\textrm{old}|$ denotes pairwise relations among Core authors in the training period.
Column $|E_\textrm{new}|$ denotes new pairwise relations among Core authors formed in the test period.}
\label{tbl:stats}
\footnotesize
\setlength\tabcolsep{1pt}
\begin{tabular}{|l|c|c|c|c|c|c|} \hline
& \multicolumn{3}{|c|}{Training Period} & \multicolumn{3}{|c|}{Core} \\ \cline{2-7}
& Authors & Articles & Collaborations & Authors & $|E_\textrm{old}|$ &  $|E_\textrm{new}|$ \\ \hline
astro-ph & 5321/5343 & 5820/5816 & 41575/41852 & 1563/1561 & 6189/6178 & 5719/5751 \\ \hline
cond-mat & 5797/5469 & 6698/6700 & 23373/19881 & 1387/1253 & 2367/1899 & 1514/1150 \\ \hline
gr-qc & 2150/2122 & 3292/3287 & 5800/5724 & 484/486 & 523/519 & 397/400 \\ \hline
hep-ph & 5482/5414 & 10277/10254 & 47788/47806 & 1786/1790 & 6629/6654 & 3281/3294 \\ \hline
hep-th & 5295/5241 & 9497/9498 & 15963/15842 & 1434/1438 & 2316/2311 & 1569/1576 \\ \hline
\end{tabular}
\end{table*}

\subsection{Link prediction in arXiv} \label{sec:arxiv}

This section replicates the experiments in \cite{liben} which are the following.
For five areas in arXiv, given the coauthors of papers published in the training period 1994-1996,
the task is to predict new pairwise coauthorship relations formed in the test period 1997-1999.
Predictions are only scored for those within \emph{Core} authors, defined as
those who have at least 3 papers in the training period and at least 3 papers in the test period;
this Core list is unknown to predictors.
Table~\ref{tbl:stats} gives statistics of the five graphs and prediction tasks.
Let $E_\textrm{new}$ be the set of new pairwise coauthorship relations among Core authors formed in the test period.
Let $E_p$ be the top $|E_\textrm{new}|$ pairs among Core authors that are predicted by a predictor,
and the score of this predictor is defined as $|E_p \cap E_\textrm{new}|/|E_\textrm{new}|$.

Table~\ref{tbl:stats} shows that our setup matches \cite{liben} closely for
four of the five benchmarks.
We focus on benchmarks astro-ph, hep-ph and hep-th,
for the following reasons.
Benchmark cond-mat differs significantly from that reported in \cite{liben},
thus is not a valid benchmark to compare against the oracle of \cite{liben}.
In gr-qc, 131 out of the 397 new relations were formed by a single project
which resulted in three papers in the test period,
with nearly identical 45-46 authors, \cite{blackhole} being one of the three.
Because the size of gr-qc is not large enough relative to this single event,
the scores of the predictors are distorted.
Thus it is not a surprise that \cite{liben} reported that the best predictor for
gr-qc is one that deletes 85-90\% of edges as a preprocessing step,
and that the same predictor delivers poor performance on the other benchmarks.

\begin{table*}[t]
\centering
\caption{Comparison of predictor accuracies on coauthorship networks.
$A$ denotes the accuracy score of a predictor.
$R$ denotes the ratio of $A$ over that of oracle of \cite{liben}.}
\label{tbl:arxiv}
\footnotesize
\setlength\tabcolsep{1pt}
\begin{tabular}{|l|l|c|c|c|c|c|c|} \hline
& & \multicolumn{2}{|c|}{astro-ph} & \multicolumn{2}{|c|}{hep-ph} & \multicolumn{2}{|c|}{hep-th} \\ \cline{3-8}
& parameters & $A$ & $R$ & $A$ & $R$ & $A$ & $R$ \\ \hline
Oracle of \cite{liben}     & varying                         & 8.55\% &         & 7.2036\% &       & 7.9407\% & \\ \hline
Oracle of Blink, graph \#1 & varying                         & 9.075\%  & 1.061 & 8.3816\% & 1.164 & 8.8592\% & 1.116 \\ \hline
Blink, graph \#1           & $b_1=0.5$, $b_2=0.4$             & 7.7461\% & 0.906 & 7.8025\% & 1.083 & 8.0306\% & 1.011 \\ \hline
Blink, graph \#2           & $b_1=0.8$, $b_2=0.6$, $\gamma=5$ & 10.264\% & {\bf{1.200}} & 9.6922\% & {\bf{1.345}} & 9.0504\% & {\bf{1.140}} \\ \hline
PPR, graph \#2             & $\alpha=0.50$                    & 8.5330\% & 0.998 & 6.7358\% & 0.935 & 7.9031\% & 0.995 \\ \hline
Modified Katz, graph \#2   & $b_1=0.5$, $b_2=0.1$, $\beta=0.1$, $\gamma=5$  & 8.4106\% & 0.984 & 8.2292\% & 1.142 & 7.8394\% & 0.987 \\ \hline
ERD, 10K samples, graph \#1 & $b_1=0.9$, $b_2=0.9$             & 8.4281\% & 0.986 & 8.1682\% & 1.134 & 7.1383\% & 0.899 \\ \hline
ERD, 10K samples, graph \#2 & $b_1=0.9$, $b_2=0.9$, $\gamma=4$ & 9.5471\% & 1.117 & 8.7473\% & 1.214 & 7.1383\% & 0.899 \\ \hline
\end{tabular}
\end{table*}

In Table~\ref{tbl:arxiv}, the oracle of \cite{liben} is the highest score for
each benchmark, by all predictors including meta-approaches;
note that no single predictor has such performance,
and PPR and Katz on uniformly weighted graphs are dominated by the oracle.
In graph \#1, each paper is modeled by a hyperedge with uniform weight.
Allowing the best $b_1$ and $b_2$ per graph leads to the oracle of Blink which easily
beats the oracle of \cite{liben};
for a single parameterization, we get the next row where Blink wins two out of three.
Such performance already puts Blink above all predictors reported in \cite{liben}.

In graph \#2, each paper is modeled as a node,
and it connects to and from each of its authors with two directed simple edges.
We provide domain knowledge through the following $f_E$ and $f_V$.
For an edge $e=(\textrm{X},\textrm{Y})$, $f_E(e) = 1 / \max ( 1 , \log_\gamma d_\textrm{X} )$,
where $d_\textrm{X}$ is the out degree of X.
For a paper node, we set $f_V$ to infinity and hence weight to 1.
For an author node, $f_V(\textrm{author}) = 1 / \max ( 1 , \log_\gamma m_\textrm{author} )$,
where $m_\textrm{author}$ is the number of coauthors of this author in the training period.
$\gamma$ is a tunable parameter
and is scanned with other parameters and reported in Table~\ref{tbl:arxiv}.

With graph \#2, the predictor scores are asymmetric.
For PPR and Katz, we experimented with max, min, sum and product of
two directional scores, and the max gives the best results and is reported.
For the Blink Model, we define symmetric
$\textrm{score}(\textrm{A,B}) = - \log ( 1-b ( \text{A},\text{B} ) \cdot b ( \text{B},\text{A} ) )$.
For ERD, the shortest-path distance in a Monte Carlo sample is defined as the shorter
between the A-to-B path and the B-to-A path.

\begin{figure}
\centering
\subfloat[astro-ph]{\includegraphics[width=1.7in]{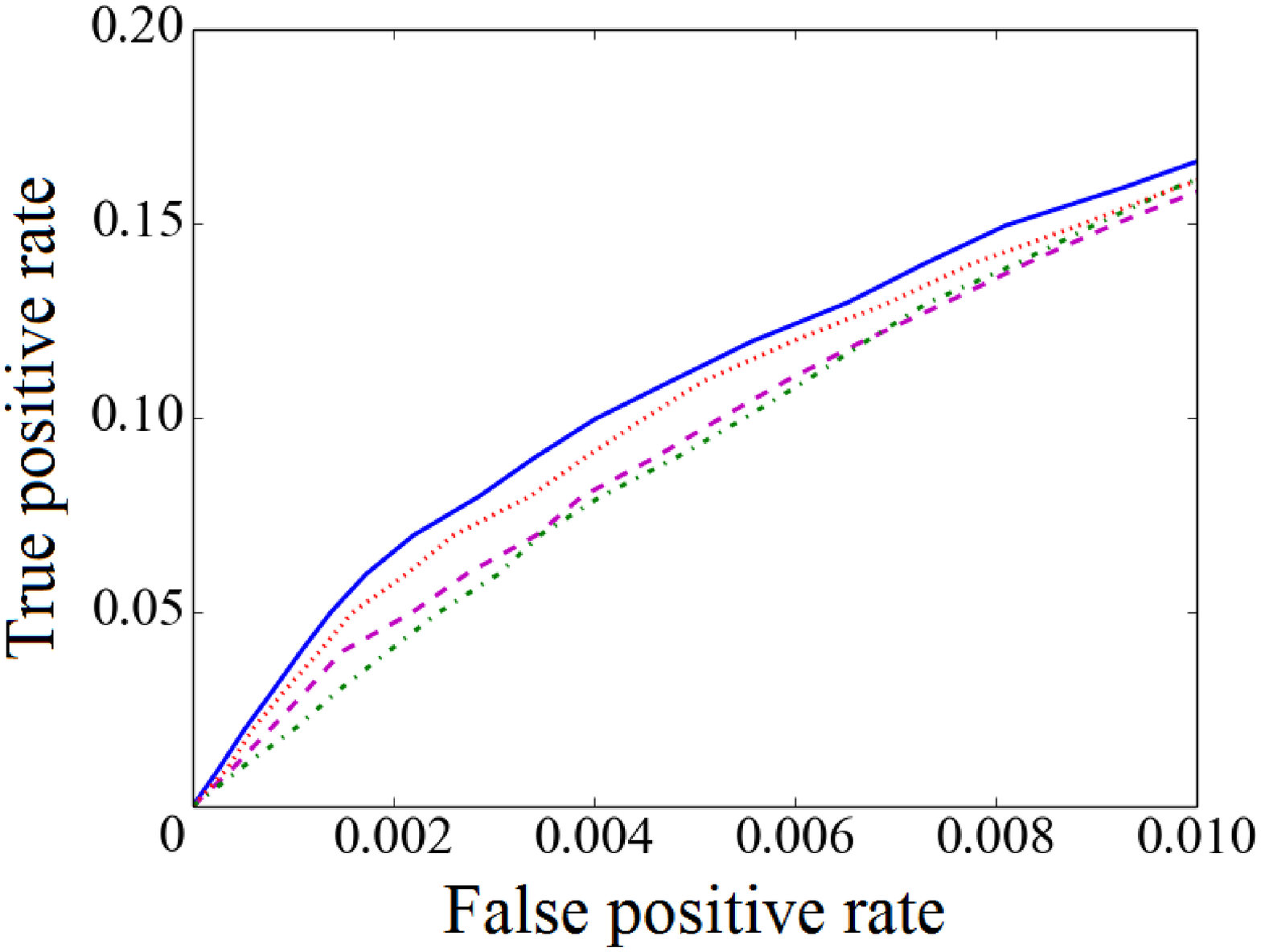}}
\subfloat[hep-ph]{\includegraphics[width=1.7in]{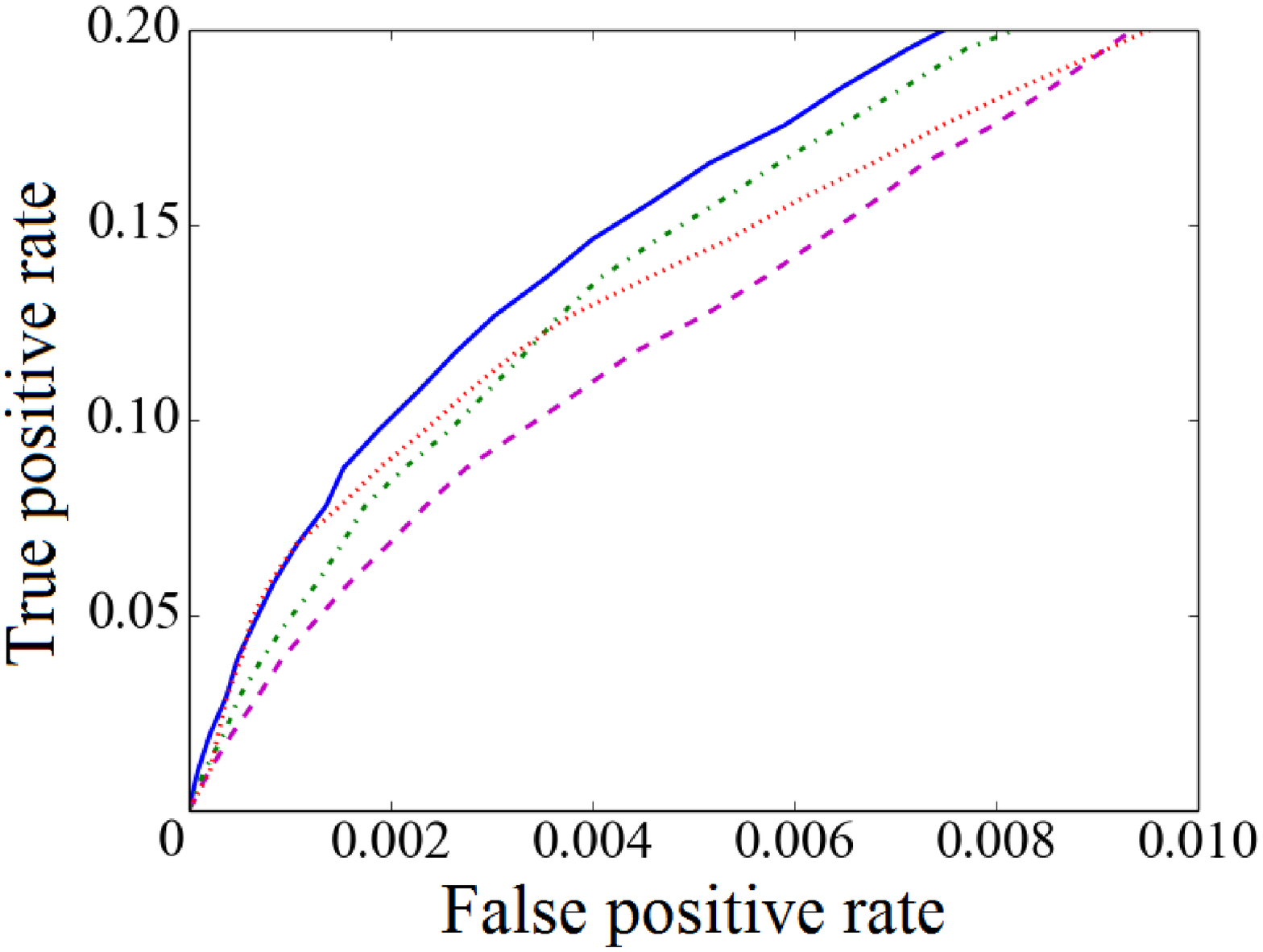}}\\
\subfloat[hep-th]{\includegraphics[width=1.7in]{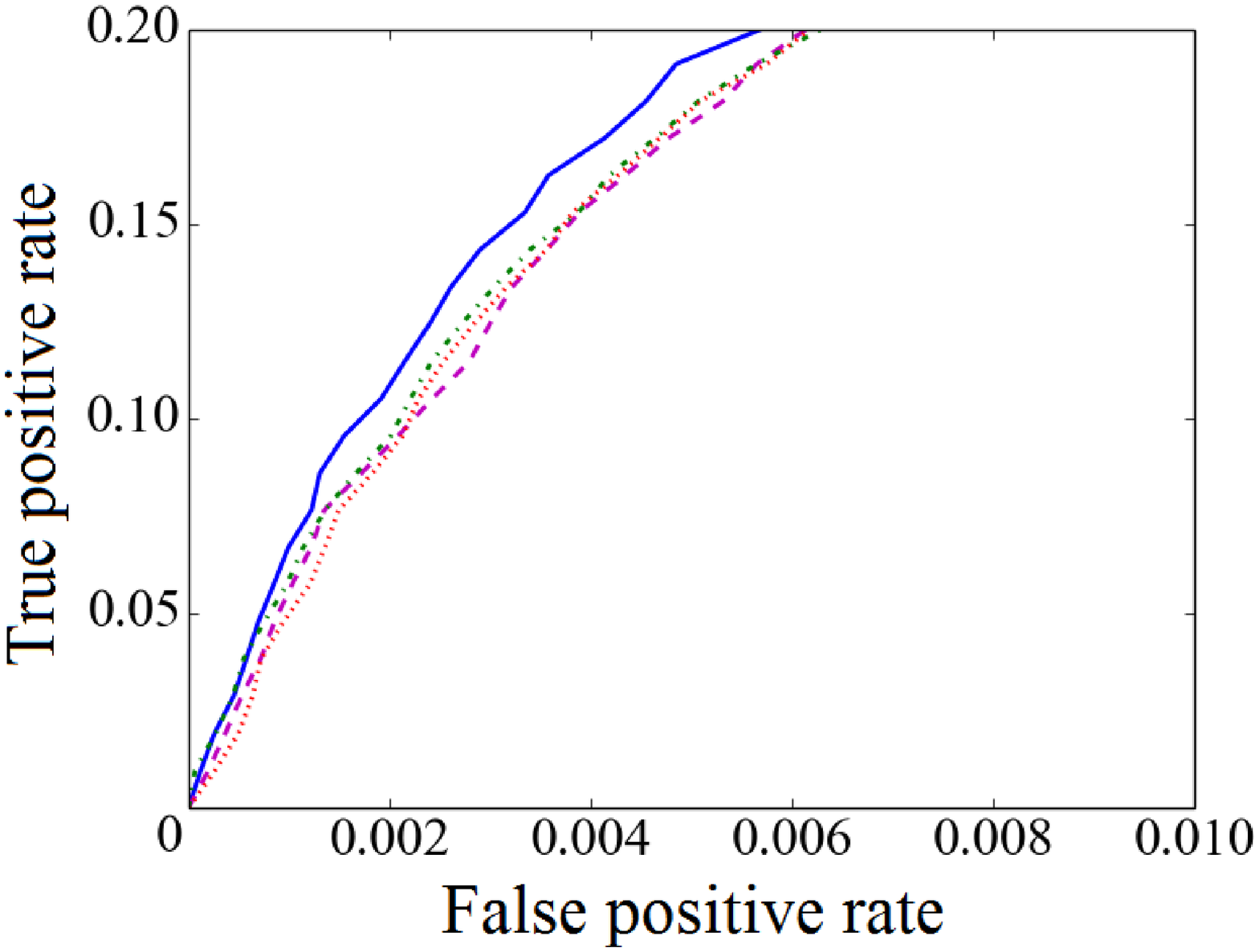}}
\subfloat{\includegraphics[width=1.7in]{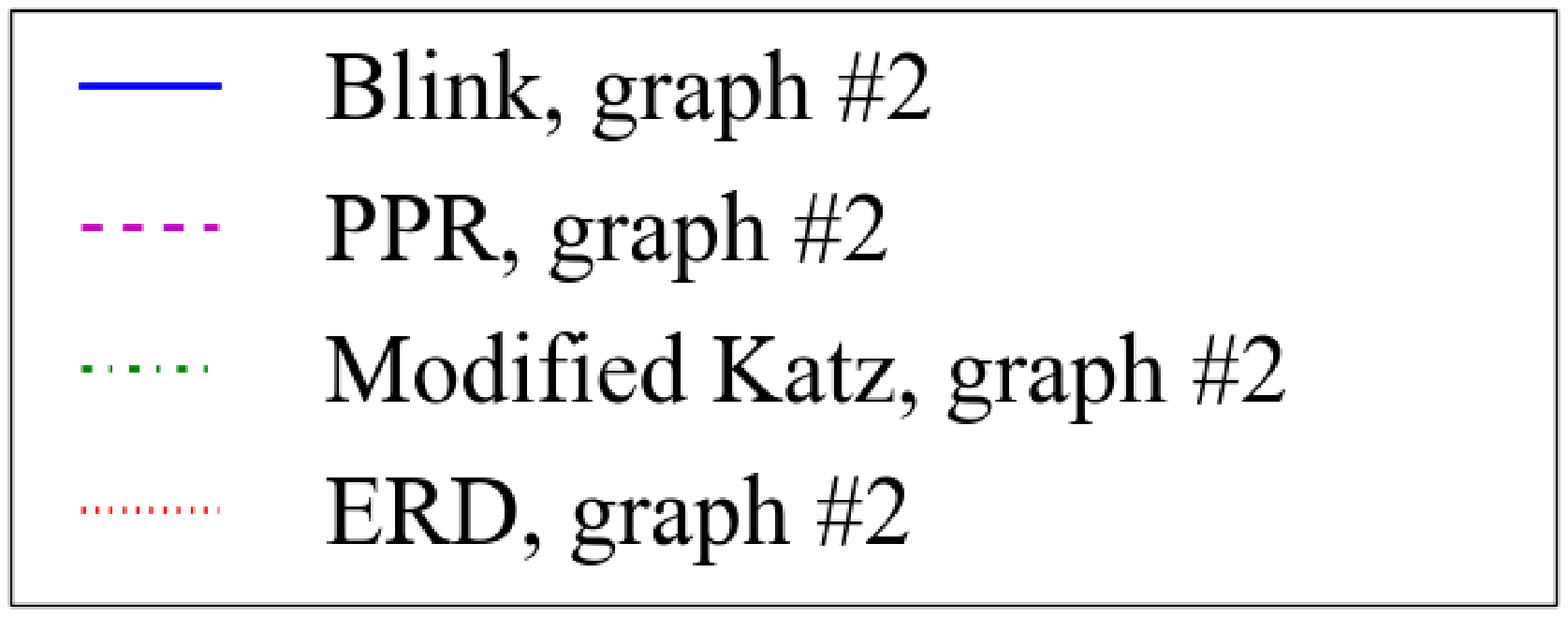}}
\caption{ROC curves on coauthorship networks.}
\label{fig:roc}
\end{figure}

Table~\ref{tbl:arxiv} demonstrates the proposed proximity measure's superior predictive power
over competitors, as expected from discussions in Section~\ref{sec:metric},
and a single parameterization of Blink Model outperforms oracle of \cite{liben} by 14--35\%.
Figure~\ref{fig:roc} shows receiver-operating-characteristic (ROC) curves.
Blink with graph \#2 clearly dominates the ROC curves.
There is not a clear runner-up among the three competitors.
Note that ERD \#2 performs well on hep-ph for early guesses at around 5\% true positive rate,
but it degrades quickly after that and becomes the worst of the four by the 20\% rate.

\begin{table*}[ht]
\centering
\caption{Statistics of Wikipedia citation networks and prediction tasks.
$n_\textrm{page}$ denotes the number of pages;
$d_{2014}$ and $d_{2015}$ denote the average number of out-going citations on a 2014/2015 page;
$d_{2014,\textrm{unique}}$ and $d_{2015,\textrm{unique}}$ denote the average number of unique out-going citations on a 2014/2015 page;
$n_\textrm{prediction}$ denotes the average number of additions to predict per task.}
\label{tbl:wikistats}
\footnotesize
\setlength\tabcolsep{1pt}
\begin{tabular}{|l|c|c|c|c|c|c|} \hline
& $n_\textrm{page}$ & $d_{2014}$ & $d_{2014,\textrm{unique}}$ & $d_{2015}$ & $d_{2015,\textrm{unique}}$ & $n_\textrm{prediction}$ \\ \hline
2014 all pages     & 4731544 & 24.66 & 23.99 &       &       & \\ \hline
2015 all pages     & 4964985 &       &       & 24.90 & 24.23 & \\ \hline
qualified tasks    & 93845   & 156.1 & 151.0 & 162.4 & 157.0 & 10.06 \\ \hline
training tasks     & 1000    & 142.2 & 137.9 & 147.5 & 143.1 & 10.14 \\ \hline
test tasks         & 1000    & 159.0 & 153.3 & 165.7 & 159.6 & 9.85 \\ \hline
trimmed test tasks & 949     & 157.6 & 151.7 & 164.3 & 157.9 & 4.63 \\ \hline
\end{tabular}
\end{table*}

\subsection{Link prediction in Wikipedia} \label{sec:wikipedia}

Our second experiment is
predicting additions of inter-wikipage citations in Wikipedia.
The rationale is that citation links reflect
Wikipedia contributors' perception of relation strength between subjects.

We obtain an inter-wikipage-citation graph from \cite{dbpedia14}
which was based on Wikipedia dumps generated in April/May 2014,
and another graph from \cite{dbpedia15} which was based on those in February/March 2015.
In both graphs, each node represents a Wikipedia page,
and each directed edge from node A to node B represents a citation on page A to page B.
The ordering of out-going citations from a page A, as they appear in the text on page A,
is known and will be utilized by some predictors.
Statistics are shown in Table~\ref{tbl:wikistats}.
4,631,780 of the 2014 nodes are mapped to 2015 nodes by exact name matching,
and another 87,368 are mapped to 2015 nodes by redirection data from \cite{dbpedia15}
which are pages that have been renamed or merged.
The remaining 12,396 of the 2014 nodes cannot be mapped:
the majority are Wikipedia pages that have been deleted,
and some are due to noise in the data collection of \cite{dbpedia14,dbpedia15}.
Such noise is a small fraction and has negligible impact to our measurements.

For each mapped node A,
we identify $S_{\textrm{A},2014}$ as the set of 2014 nodes that page A cites in the 2014 graph and that remain in the 2015 graph,
$S_{\textrm{A},2015}$ as the set of 2014 nodes that page A cites in the 2015 graph,
and $X_{\textrm{A},2014}$ as the set of 2014 nodes that cite page A.
If page A satisfies the condition that
$5 \leq |\left( S_{\textrm{A},2015} \backslash S_{\textrm{A},2014} \right) \backslash X_{\textrm{A},2014} | \leq |S_{\textrm{A},2014}|\cdot20\%$,
we consider page A as a qualified prediction task. 
The rationale behind the size limits is to choose test pages that have undergone thoughtful edits,
and their 2014 page contents were already relatively mature;
the rationale for excluding in-coming neighbors $X_{\textrm{A},2014}$ is to make
the tasks more challenging, since simple techniques like heavily weighting in-coming edges have no effect.
Statistics are shown in Table~\ref{tbl:wikistats}.
The number of qualified tasks is large,
and we randomly sample a 1000-task training set and a 1000-task test set.

Tasks vary in difficulty. If edits were to make a page more complete,
the new links are often reasonably predictable and some are obvious.
However, if edits were driven by a recent event, the new links are next to impossible to predict.
We form a trimmed test set by removing from the test set
targets that are too easy or too difficult, utilizing the outputs of four
best-performing predictors on the test set: Adamic/Adar, Blink Model \#2, Personalized PageRank \#2,
and Modified Katz \#2.
A new link is removed if it ranks less than 20 by all four predictors or if it ranks more than 1000 by all four.
The removed prediction targets are excluded from predictor outputs during mean-average-precision evaluation.
The results are listed in the last row of Table~\ref{tbl:wikistats}
and the last two columns of Table~\ref{tbl:wiki}.

\begin{table*}[ht]
\centering
\caption{Comparison of predictor accuracies on additions of inter-wikipage citations in Wikipedia.
Each predictor uses its best parameters selected based on the training set.
$R$ denotes the ratio of MAP of a predictor over MAP of Adamic/Adar.}
\label{tbl:wiki}
\footnotesize
\setlength\tabcolsep{1pt}
\begin{tabular}{|l|l|c|c|c|c|c|c|} \hline
& & \multicolumn{2}{|c|}{training} & \multicolumn{2}{|c|}{test} & \multicolumn{2}{|c|}{trimmed test} \\ \cline{3-8}
                & parameters & MAP & $R$ & MAP & $R$ & MAP & $R$ \\ \hline
Adamic/Adar     &                                    & 0.0291 &       & 0.0281 &       & 0.0163 & \\ \hline
Blink, graph \#1 & $b_1=0.5$, $b_2=0.1$               & 0.0295 & 1.014 & 0.0263 & 0.937 & 0.0166 & 1.019 \\ \hline
Blink, graph \#2 & $b_1=0.8$, $b_2=0.8$, $\gamma=10$  & 0.0362 & 1.244 & 0.0362 & {\bf{1.289}} & 0.0233 & {\bf{1.428}} \\ \hline
PPR, graph \#1   & $\alpha=0.5$                       & 0.0299 & 1.029 & 0.0291 & 1.038 & 0.0186 & 1.140 \\ \hline
PPR, graph \#2   & $\alpha=0.2$, $\gamma=500$         & 0.0321 & 1.104 & 0.0309 & 1.100 & 0.0206 & 1.263 \\ \hline
Modified Katz, graph \#1 & $\beta=5\textrm{E-}6$      & 0.0269 & 0.925 & 0.0241 & 0.860 & 0.0151 & 0.924 \\ \hline
Modified Katz, graph \#2 & $b_1=0.8$, $b_2=0.8$, $\beta=0.1$, $\gamma=10$ & 0.0341 & 1.173 & 0.0328 & 1.170 & 0.0198 & 1.213 \\ \hline
ERD, 100 samples, graph \#1 & $b_1=0.4$, $b_2=0.9$                 & 0.0266 & 0.914 & 0.0233 & 0.830 & 0.0162 & 0.996 \\ \hline
ERD, 100 samples, graph \#2 & $b_1=0.9$, $b_2=0.9$, $\gamma=10$    & 0.0238 & 0.817 & 0.0218 & 0.778 & 0.0154 & 0.944 \\ \hline
\end{tabular}
\end{table*}

In Table~\ref{tbl:wiki}, mean average precision (MAP) is the accuracy score.
Unlike in Section~\ref{sec:arxiv}, the relations to predict
are asymmetric (page A adds a citation to page B)
and hence all predictors use their one-directional A-to-B score as is.
When multiple node B's have the same score, we use their in-coming degrees as a tie breaker:
a node with higher in-coming degree is ranked first.
Adamic/Adar is used as a reference, as it represents what can be achieved through good-quality local analysis.
In graph \#2, we provide domain knowledge through the following $f_E$ and $f_V$.
For an edge from node X to node Y and that represents the $i^\textrm{th}$ citation link on page X:
\begin{equation}
f_E(\textrm{edge}) = \frac{\delta_\textrm{Y,X}}{\max \left( 1 , \log_\gamma i \right) \cdot \max \left( 1 , \log_\gamma{d_\textrm{Y,in}} \right)}
\end{equation}
where $\delta_\textrm{Y,X}$ is 2 if edge exists from Y to X, and 1 otherwise.
$\gamma$ is again a tunable parameter.
The above scheme gives higher weight to a citation link if it is located at an earlier location on a page,
or if it points to a less-cited page, or if a returning citation exists.
Our $f_V$ function is a direct adaptation of Adamic/Adar's (\ref{eq:adamicadar}):
$f_V(\textrm{node}) = 1/( \log{d_\textrm{node,in}}+\log{d_\textrm{node,out}} )$.
To get best results, PPR uses (\ref{eq:linearweight}) while Katz and ERD use (\ref{eq:weight}).
A remarkable observation on Katz with graph \#2 is that its best
performance happens when (\ref{eq:katz}) is almost divergent:
with $b_1=0.8$, $b_2=0.8$ and $\gamma=10$, the divergence limit for $\beta$ is 0.1075.
For ERD, the reduction in sample numbers from Section~\ref{sec:arxiv} is because the Wikipedia graph is much larger and denser.
Blink Model with graph \#2 is clearly the best performer in Table~\ref{tbl:wiki}.

\begin{figure}[ht]
\centering
\subfloat[Test set]{\includegraphics[width=3in]{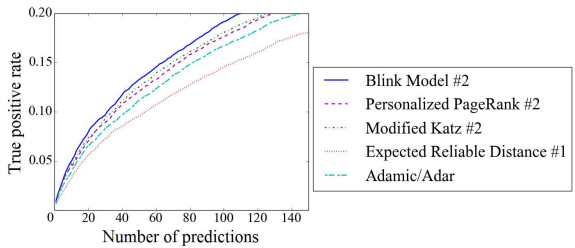}}\\
\subfloat[Trimmed test set]{\includegraphics[width=3in]{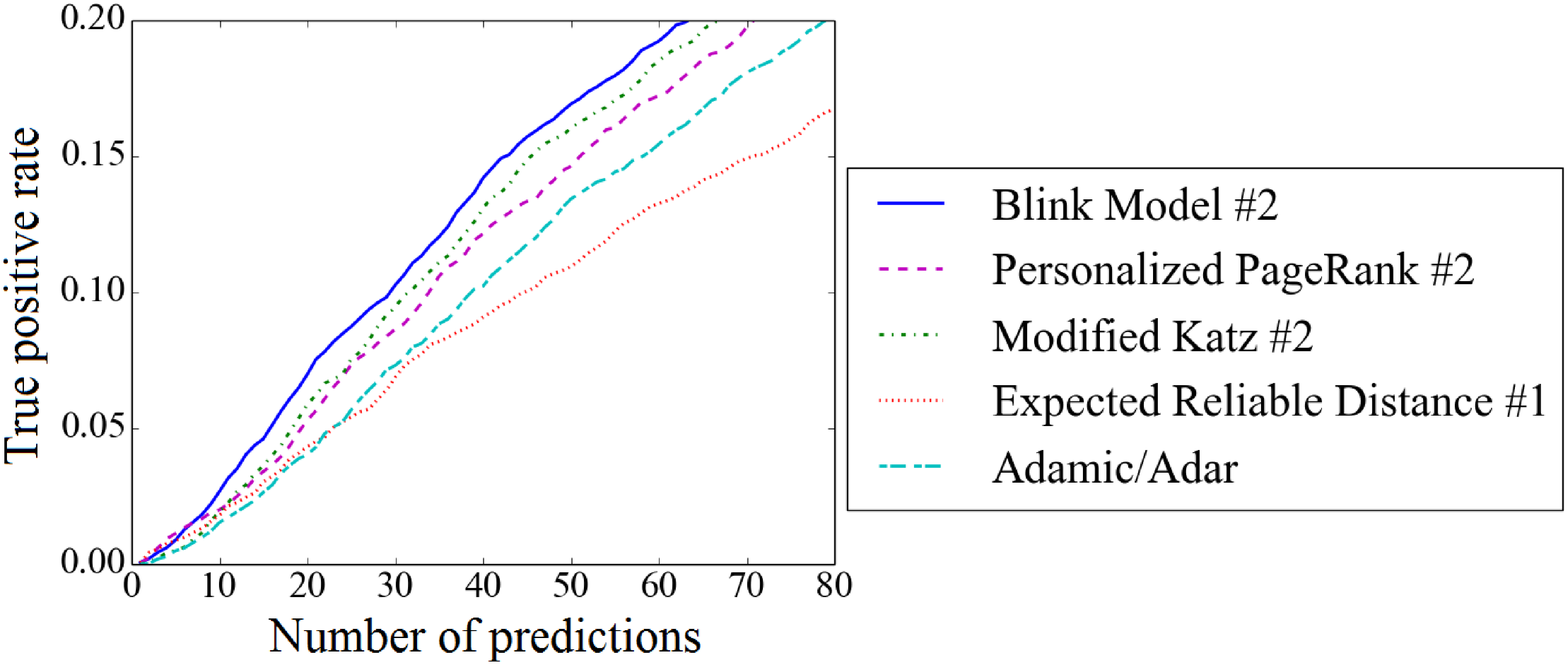}}
\caption{Accuracy curves on Wikipedia citation network.}
\label{fig:curve}
\end{figure}

Figure~\ref{fig:curve} shows a more detailed comparison by plotting
true positive rate as a function of the number of predictions made.
Blink Model \#2 clearly dominates the curves, and for example needs 20 fewer predictions to reach
a 20\% true positive rate on the test set than the closest competitor.

\section{Conclusions}

This manuscript proposes the Blink Model graph proximity measure.
We demonstrate that it matches human intuition better than others,
develop an approximation algorithm,
and empirically verify its predictive accuracy.

\bibliographystyle{IEEEtran}
\bibliography{prediction}

\end{document}